\documentclass[aps,prl,twocolumn,showpacs,superscriptaddress,groupedaddress]{revtex4-1}
\usepackage{graphicx}
\usepackage{graphics}

\usepackage{booktabs}

\usepackage{dcolumn}   
\usepackage{bm}        
\usepackage{amssymb}   

\begin{document}

\title[dSphs in multistate SFDM haloes]{Dwarf galaxies in multistate Scalar Field Dark Matter haloes}
\author{L. A.~Martinez-Medina}
\email{lmedina@fis.cinvestav.mx}
\affiliation{Departamento de F\'isica, Centro de Investigaci\'on y de Estudios Avanzados del IPN, A.P. 14-740, 07000 M\'exico D.F., M\'exico.}
\author{V. H. Robles}
\email{vrobles@fis.cinvestav.mx}
\affiliation{Departamento de F\'isica, Centro de Investigaci\'on y de Estudios Avanzados del IPN, A.P. 14-740, 07000 M\'exico D.F., M\'exico.}
\author{T.~Matos}
\email{tmatos@fis.cinvestav.mx}
\affiliation{Departamento de F\'isica, Centro de Investigaci\'on y de Estudios Avanzados del IPN, A.P. 14-740, 07000 M\'exico D.F., M\'exico.}



\label{firstpage}

\begin{abstract} 

We analyse the velocity dispersion for eight of the Milky Way dwarf spheroidal satellites in the context of finite 
temperature scalar field dark mater. In this model the finite temperature allows the scalar field to be in configurations 
that possess excited states, a feature that has proved to be necessary in order to explain the asymptotic rotational velocities 
found in low surface brightness (LSB) galaxies. In this work we show that excited states are not only important in large galaxies but 
also have visible effects in dwarf spheroidals.  
Additionally, we stress that contrary to previous works where the scalar field dark matter haloes are consider to be purely 
Bose-Einstein condensates, the inclusion of excited states in these halo configurations provides a consistent framework 
capable of describing LSBs and dwarf galaxies of different sizes without arriving to contradictions within the scalar field 
dark matter model. Using this new framework we find that the addition of excited states accounts very well for the raise 
in the velocity dispersion in Milky Way dwarf spheroidal galaxies improving the fit compared to the one obtained assuming  
all the DM to be in the form of a Bose Einstein Condensate.

\end{abstract}

\maketitle


\section{Introduction}
\label{sec:Intro}

The higher precision in observations at galactic scale reveals important deviations from the predictions of the standard cold dark matter model (CDM), such as the too big to fail \citep{Boylan2011, Rashkov2012, Tollerud2012} and a strong relation with the known cusp-core problem \citep{dB01, K03, Gentile2004, G2006, Battaglia2008, Walter2011, O11, R13a, A12}.

This discrepancies have motivated other alternative dark matter (DM) models like the scalar field dark matter (SFDM) \citep{G00,M01, Bray2013, R13a, Bray2014} or self interacting dark matter \citep{Spergel2000, Zavala2009}  that produce inner constant density distributions (usually referred as core profiles) through some mechanism intrinsic to dark matter properties rather than extreme baryonic processes that are still not fully understood. 

The SFDM model proposes that the dark matter is a real scalar field minimally coupled to gravity with a small mass of $m \sim10^{-21}eV/c^2$. At zero temperature all particles in the system condense to the same quantum ground state forming a Bose-Einstein condensate (BEC). The SFDM model has being extensively study in other works and proved to be in agreement with large scale observations and with rotation curves of low mass galaxies \citep{Bohmer2007, Rodriguez2010, Suarez2011, Lora2012, Lora2014, Harko2011, Chavanis2012, R12, R13a, Schive2014, MartinezMedina2014}. 

When the dark matter behaves as a Bose-Einstein condensate (BEC), a universal mass density profile for the localized, static, spherically symmetric scalar field configurations emerges with the following analytic form:

\begin{equation}
\label{eq:density0}
\rho(r) = \rho^0_0\frac{\sin(\pi r/r^0_{max})}{(\pi r/r^0_{max})},
\end{equation}

where $\rho^0_0$ is the central density of the condensate, $r^0_{max} = \sqrt{\pi^2\Lambda/2}(\hbar/mc)$ is considered as a configuration radius where $\rho(r^0_{max}) = 0$ and 
therefore $\rho(r) = 0$ for $r\geq r^0_{max}$ , $m$ is the mass of the scalar field, $\Lambda= \lambda m^2_{Planck}/4\pi m^2$, and $\lambda$ is the adimensional parameter that determines the two body interactions of the field. 

The viability of these configurations as dark matter haloes has been tested in several works, in particular fitting rotation curves (RCs) in low surface brightness (LSB) galaxies \citep{Bohmer2007, R12, R13a}. Such  analysis is suitable for those rotating spiral galaxies where the circular velocity relates directly to the enclosed mass at a radius $r$, but it is not the most suitable approach for systems where ordered rotation is dynamically negligible, this is the case of several dwarf spheroidals (dSphs) in the Milky Way. These systems are among the least luminous and more DM dominated structures in the universe with stellar populations assumed to be in dynamic equilibrium and pressure supported. Due to their high mass to light ratios, they are good candidates to test predictions of DM models such as the halo shape and density distribution using the line-of-sight velocity dispersion.

Recently \citet{Diez2014} assume the profile in equation (\ref{eq:density0}) to represent a typical dSph DM halo, they conclude that the best fits to the velocity dispersion data for the eight brightest dSph satellites of the Milky Way suggest either a typical scale radius for a BEC halo of $r^0_{max} \sim 1$kpc, or an unnatural preference for radial orbits in case haloes have larger scale radii and are described by the same profile, these results lead the authors to point out an apparent incompatibility with the scales reported in previous studies of LSB galaxies where typical scale radii are $\sim 5-7$~kpc  as opposed to $\sim$1~kpc commonly found in dwarfs.

However, there are some factors that need to be considered when interpreting their results, one of them is that the scale radii of the condensate in different haloes is expected to have some scatter from the typical value 1 kpc.  The outer structure of a given dwarf is affected by several factors, like its formation history, its evolution as a satellite where tidal forces may play a significant role in shaping the outer structures, specially for those closest to their host \citep{Hodge1966, Mayer2001, Munoz2006, Penarrubia2010, R14}, among others. It s then likely that environmental effects change the structure and kinematics of a given satellite and thus produce some scatter in their velocity dispersion profiles and scale radii.

On the other hand, we will stress that accounting for the possibility of excited state configurations for the scalar field avoids any tension and therefore eliminates the need to have only one relevant scale length for the ground (condensed) state in all galaxies, the idea to include more than the ground state has been explored in other contexts too \citep{Urena2010, Bernal2010, R13b}.

In order to show this, we will not restrict the description of the SFDM halo by the mass density profile in eq (\ref{eq:density0}), 
derived under the assumption that \textit{all} DM particles are in a condensed state at temperature $T=0$, instead we consider one 
scenario that allows excited states of the SF when dark matter temperature corrections to first order are taken into account \cite{R13a}. 
This scenario is capable of reproducing rotation curve data that extends to large radii \citep{R13a, MartinezMedina2014}, a feature that 
is not present when the halo is only in its ground state. In this paper we will explore the consequences of excited states in dSphs in 
the same scenario that already explains the flatness of the RCs in larger galaxies. 

In \cite{R13a} they found an analytic form for the density distribution of a static spherically symmetric SFDM halo in the state $j$ given by 

\begin{equation}
\label{eq:density}
\rho_j(r) = \rho^j_0\frac{\sin^2(k_jr)}{(k_jr)^2},
\end{equation}
here $\rho^j_0$ is the central density of DM in state $j$, $k_j := \pi j/R_j$, $R_j$ is a cut-off radius such that  $\rho(r) = 0$  for all $r \geq R_j$,  and $j$ is a positive integer that determines the minimum number of excited states required to model a galaxy DM halo, notice that the ground state corresponds to $j$=1 and it has no oscillations in the density profile.

We notice that the ground state $j$=1 from eq. (\ref{eq:density}) is not 
the same for the BEC density profile from eq. (\ref{eq:density0}). Equation (\ref{eq:density}) is a solution of the equation of motion for a scalar field perturbation when the self-interaction parameter $\lambda$ is small, which is true according to current bounds on the self-interaction strength obtained in \citep{Rindler2014} for the small mass of the scalar field proposed in the SFDM model.
For such small values of $\lambda$, our analytical solutions are expected to be more similar
to the complete numerical solutions obtained without self-interactions in the field. Meanwhile, equation (\ref{eq:density0}) comes from the Thomas-Fermi 
approximation \citep{Bohmer2007}, this considers that the whole system is in the ground state at $T=0$, then in this case the self-interaction term is dominant and only the ground state solution has physical interpretation. As is expected, the self-interactions widen the distribution of DM and creates the difference of equations \ref{eq:density0} and \ref{eq:density} for the ground state.

From eq. (\ref{eq:density}) the mass enclosed within a radius $r$, for the profile given in equation (\ref{eq:density}), is given by 

\begin{equation}
\label{eq:mass}
M_j(r) = \frac{2\pi\rho_0}{k_j^2}\left[ r - \frac{\sin(2k_jr)}{2k_j} \right]
\end{equation}

In this article we will explore the consequences and effects of considering excited states in the SF to model the MW dSphs and discuss the differences between this and the previous full-condensate analysis, we begin in Section 2 writing the velocity dispersion equations for our analysis, in Section 3 we present the results and section 4 is devoted to conclusions.

\section{Velocity dispersion model}
\label{sec:Jeans}

We assume that the stellar systems of the dSphs are pressure supported and in dynamic equilibrium, we follow the procedure of \citet{Walker2009} to find the relationship between the mass distribution of the DM halo and the stellar distribution. This is given by the Jeans equation

\begin{equation}
\label{eq:jeans}
\frac{1}{\nu}\frac{d}{dr}(\nu \langle v_r^2\rangle) + 2\frac{\beta\langle v_r^2\rangle}{r} = -\frac{GM(r)}{r^2},
\end{equation}
where $\nu(r)$,$\langle v_r^2\rangle$, and $\beta(r)$ describe the 3-dimensional density, radial velocity dispersion, and orbital anisotropy, respectively, of the stellar component.

The parameter $\beta$ quantifies the degree of radial stellar anisotropy and there is no preference for either radially, $\beta > 0$, or tangentially, $\beta < 0$, biased systems. 

For circular orbits $\beta = \infty$ ,$\langle v_r^2\rangle = 0$ ; if $\langle v_r^2\rangle = \langle v_{\theta}^2\rangle$, $\beta = 0$ (isotropic orbits); and for radial orbits $\beta = 1$, $\langle v_{\theta}^2\rangle = 0$. Although $\beta$ can take all this values, we restrict the anisotropy to be in the range $-0.6\leq \beta \leq 0.3$ for the most realistic scenarios \citep{Walker2009}, and consistent with other estimates using also the dSphs \citep{Lokas2009}.

In the simplest case the orbital anisotropy is independent of $r$ ( $\beta = const$),  the solution of the Jeans equation relates the projection of the velocity dispersion along the line-of-sight, $\sigma_{los}^2(R)$, and the mass profile $M(r)$, to the stellar density $I(R)$ \citep{Binney87} through 

\begin{equation}
\label{eq:slos}
\sigma_{los}^2 = \frac{2G}{I(R)} \int_R^{\infty}dr\prime \nu(r\prime)M(r\prime)(r\prime)^{2\beta-2}F(\beta,R,r\prime),
\end{equation}
with
\begin{equation}
\label{eq:F}
F(\beta,R,r\prime) \equiv \int_R^{r\prime}dr \left( 1 - \beta\frac{R^2}{r^2} \right)\frac{r^{-2\beta+1}}{\sqrt{r^2-R^2}},
\end{equation}
and $R$ the projected radius. 

We adopt an analytic profile for the projected stellar density $I(R)$. As in other works \citep{Walker2010, Salucci2011}, we consider a Plummer profile for the stellar density with the projected half-light radius, $r_{half}$, as the only shape parameter, 

\begin{equation}
\label{eq:I}
I(R) = \frac{L}{\pi r_{half}^2}\frac{1}{[1+(R/r_{half})^2]^2},
\end{equation}
where $L$ is the total luminosity.
With the projected stellar density known, one can recover the 3-dimensional stellar density \citep{Binney87}

\begin{equation}
\label{eq:nuintegral}
\nu(r) = -\frac{1}{\pi}\int_r^{\infty}\frac{dI}{dR}\frac{dR}{\sqrt{R^2-r^2}}.
\end{equation}
substituting eq. (\ref{eq:I})  in eq. (\ref{eq:nuintegral}) for the Plummer profile we have
\begin{equation}
\label{eq:nu}
\nu(r) = \frac{3L}{4\pi r_{half}^3}\frac{1}{[1 + (r/r_{half})^2]^{5/2}}.
\end{equation}

In the next section we use eqs. (\ref{eq:slos}) and (\ref{eq:F}) to find the halo parameters that best reproduce the velocity dispersion data. We have three free parameters per galaxy: the scale radius $\sim 1/k_j$, the density, and the orbital anisotropy $\beta$.
For the stellar component we take $r_{half}$ from \citep{Walker2009}.

\section{Results}
\label{sec:results}

We first model the dSphs assuming the base state is enough, this means setting $j$=1 and thus eq.(\ref{eq:density}) 
has no oscillations, the situation is the analogue to that in previous studies where eq. (\ref{eq:density0}) is used to fit dwarf galaxies but 
now the pure condensate enters just as a particular case of eq.(\ref{eq:density}), the latter accommodating excited field configurations 
that are not reproduced with eq.(\ref{eq:density0}).

\begin{figure*}
\begin{center}
\includegraphics[width=18cm]{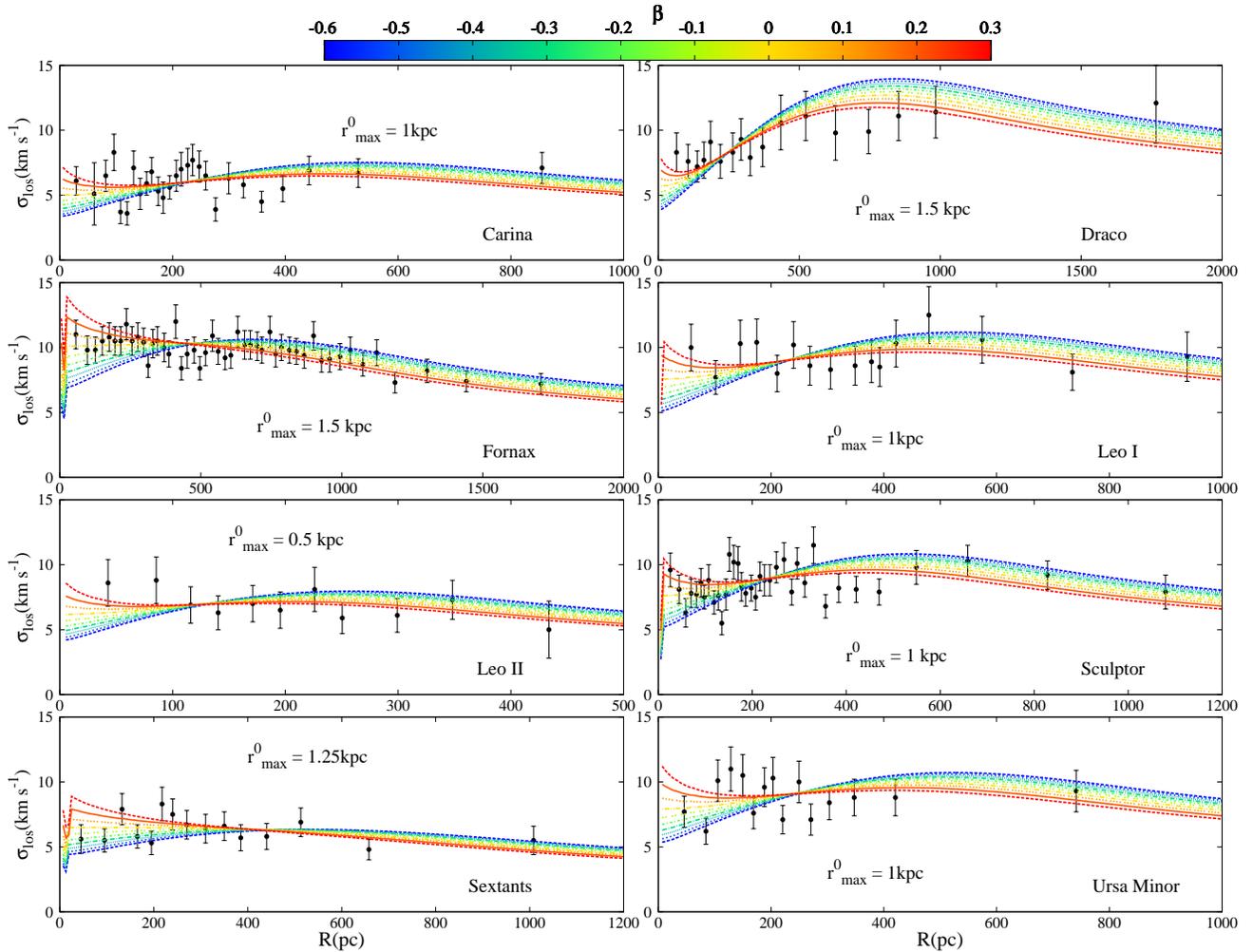}
\end{center}
\caption{Projected velocity dispersion profiles corresponding to the profiles calculated with the BEC model (eq.\ref{eq:density0})
for a range of realistic values of the orbital anisotropy. Also shown in each panel are the scale radii of the condensed state, $r^0_{max}$, that best fit the data.}
\label{fig:dispersion1}
\end{figure*}

From Figure \ref{fig:dispersion1} we observe that the data is well reproduced by the curves obtained from eq. (\ref{eq:slos}) and 
eq.(\ref{eq:density0}) restricting to realistic values of the orbital anisotropy, $-0.6\leq \beta \leq 0.3$.
In order to have the best fits when the DM halo is composed only of condensed particles, the scale radius $R_1$ of the haloes should be small, in all cases less than $2$~kpc, this result was also pointed out in \citet{Diez2014} when the DM halo is described by 
eq.(\ref{eq:density0}) and also in \citet{Walker2009} if dSph haloes have a core density.

If eq.(\ref{eq:density0}) is assumed to model dwarf haloes it is also possible to obtained good fits using larger scale radii but 
only by choosing bigger and often extreme anisotropy values \citep{Diez2014}.
Thus, if dSphs are embedded in fully condensed scalar field haloes, then they are more likely to have small scale radii ($\leq$2 kpc). 

On the other hand, if the BEC profile (eq.(\ref{eq:density0})) is used to fit rotation curves in more extended DM dominated galaxies, 
such as low surface brightness galaxies, the scale radii of these condensed haloes are found to be larger than those found 
in dSphs, on the order of $\sim$5 kpc, this then leads to a problem of changing the BEC scale radius, $r^0_{max}$, which should 
remain constant in this scenario. This issue is present as long as we want to describe the dynamics of DM dominated galaxies 
of different sizes with only the base state.

Fortunately, when we allow excited states to be present in the scalar field haloes the above ``scale issue'' can be solved and 
at the same time leaves the BEC description valid for the inner parts of the haloes.
Indeed, we see in Fig. 2 that when the density distribution changes to eq. (\ref{eq:density}) it produces dispersion velocity fits 
in dSphs that are almost indistinguishable to those using eq. (\ref{eq:density0}) within $\sim 500 pc$, the region where the 
condensed state best fits the data.
The latter implies that within this radius the dSphs are also well reproduced using eq.(\ref{eq:density}) with only the 
condensed state($j=1$). We notice too that even if the ground state dominates here, the addition of one more state produces a 
small difference with respect to the fully condensed halo fit.  

The relevance of adding excited states is mostly seen in larger galaxies as they play an important role to 
flatten the rotation curves and reach agreement with HI observations. 
Notice that the halo size depends on the number of excited states that are required to fit the outermost observations, 
therefore there is no longer need to have one scale for all galaxies, i.e., galaxies of different sizes may be embedded in 
haloes with different states. 
We also expect that dSphs of similar sizes and properties are described with a single state, in fact this is shown 
in Fig. 2 for the dSphs within $500 pc$, notice that the corrections due to the first excited state are small but not zero,  
we must take into consideration that the larger the galaxy is, the more relevant are the contributions of other states, this 
is partly the reason why dwarfs are well fitted by the base state.  

Considering that data are limited in the outer regions of the brightest dSphs of the Milky Way, it is reasonable to consider that 
the first state that could contribute to modify the dispersion profile of a condensed halo will be due to the first excited state ($j=2$). 
Therefore, as a first correction we consider configurations with both base and first states for the dSph DM haloes and fit the 
data including also the outermost measured values.

\begin{figure*}
\begin{center}
\includegraphics[width=18cm]{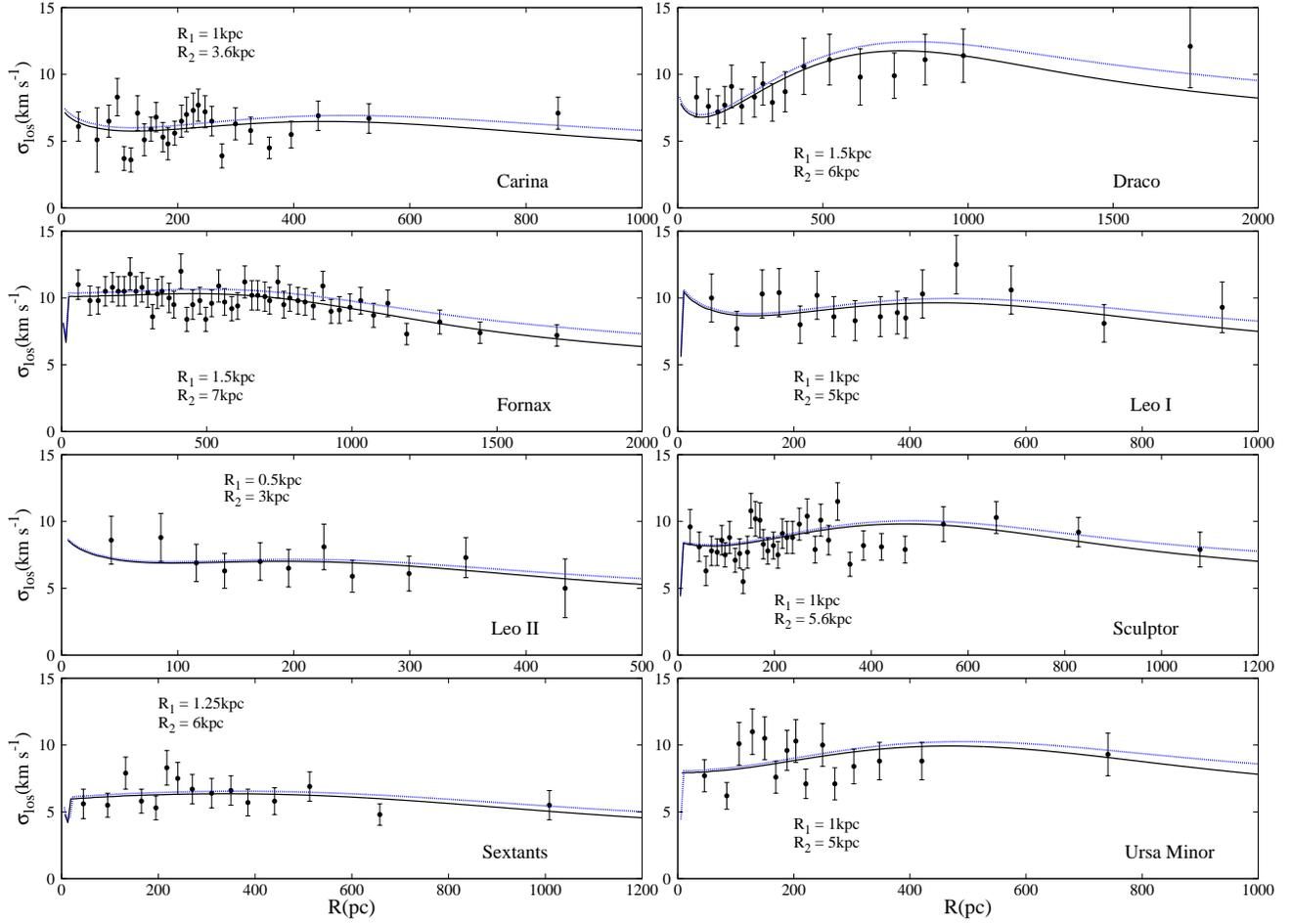}
\end{center}
\caption{Projected velocity dispersion profiles for the eight brightest dwarf spheroidal satellites of the Milky Way \citep{Walker2009}. 
The solid (black) lines correspond to the profiles calculated with the anisotropy parameter that best fits the data in Fig. \ref{fig:dispersion1} with the BEC model(eq.\ref{eq:density0}), 
and the dotted (blue) lines show the temperature corrected profile (eq.\ref{eq:density}) that includes the sum of the first excited state and the ground state.
The effects of the first state are more pronounced for the galaxies with measurements in $r>$1 kpc, these states have $R_2 > R_1$
implying that dwarf dark matter haloes extend to at least $R_2 \approx 5$ kpc, however the dominant component of the mixed state within $500$ pc 
remains to be the ground state, this region is also where the BEC profile provides a good description.}
\label{fig:dispersion2}
\end{figure*}

Figure \ref{fig:dispersion2} shows both, the profiles obtained assuming the BEC profile for the value of $\beta$ that best fit the data,
and the fit to the velocity dispersion obtained from eq. (\ref{eq:slos}) when we 
assume a configuration given by the sum of the condensed state and the first excited state for each of the eight brightest dwarf 
spheroidal satellites of the Milky Way.  
In the latter fits, for the base state we used the same value of $R_1$ found in Fig. 1 so that we could compare with the effect of 
adding the first excited state, for the first state, we search for the minimum value of $R_2$ that maintains or raises the outer 
dispersion profile such that it includes the farthest value and that keeps the profile flat for larger radii.

In Table I we summarize the values used for the fits in Fig. 2, it shows the values 
of the central densities and scale radii for each state as well as the anisotropy parameter that best fit the data in Fig. 1.
We notice that in all cases $R_1 \sim 1$kpc but we remark that $R_1$ can vary from galaxy to galaxy as this is a fitting 
parameter that represents the region where the base state dominates and does not necessarily determine the size of the halo, hence 
it is not expected to be a constant for galaxies of different sizes as opposed to the what is implied in the BEC model.

\begin{table*}
\caption{Parameters for the condensate state and the 1st excited state that together form the potential model for each dSph (blue lines in Fig \ref{fig:dispersion2}).}
\centering
\begin{tabular}{c c c c c c}
\hline \hline
dSph & $\rho^{BEC}_0$ (10$^{-2}$M$_{\odot}$pc$^{-3}$) & R$_{1}$ (kpc) & $\rho^{1st-state}_0$ (10$^{-2}$M$_{\odot}$pc$^{-3}$) & R$_{2}$ (kpc) & $\beta$ \\
\hline
Carina     & 6.60     &  1    & 0.45   &  3.6  & 0.3   \\
Draco      & 10.6     &  1.5  & 0.51   &  6    & 0.3   \\
Fornax     & 5.09     &  1.5  & 0.16   &  7    & 0     \\
Leo I      & 14.5     &  1    & 0.37   &  5    & 0.3   \\
Leo II     & 28.5     &  0.5  & 0.46   &  3    & 0.3   \\
Sculptor   & 13.5     &  1    & 0.24   &  5.6  & 0.1   \\
Sextants   & 2.42     &  1.25 & 0.07   &  6    & -0.1  \\
Ursa Minor & 13.1     &  1    & 0.34   &  5    & 0     \\
\hline
\end{tabular}
\label{tab:fit}
\end{table*}

In Fig. \ref{fig:dispersion2} we show that the inclusion of one excited state in a dwarf SFDM halo has two main 
consequences on the computed velocity dispersion. 
First, it is possible to have SF configurations that fit the data with realistic anisotropy values 
and large scale radius, this scale is possible due to the presence of the excited state that extends to 
outer radii while the condensate continues to be the dominant contribution within the typical radius $\sim 1$ kpc. 
Moreover, the addition of excited states avoids the need to choose different scales for the BEC component in LSBs and dwarf 
galaxies and finish with contradictions within the model, therefore this extension offers a self consistent framework that 
can describe a wide range of galaxies.  

Second, even though dwarfs are well fitted by a BEC halo within $1$ kpc we see that albeit small, there are visible effects 
due to the presence of higher states at these radii, but the corrections are always more pronounced for $r \geq$ 1 kpc as shown in 
Fig. 2 for Draco, Fornax and Carina. 
Unfortunately, the lack of data for $r>1$~kpc in most dSphs makes it difficult to notice the effects that higher 
states have in the fits of dispersion profiles, even with this limitation we have shown that until now if we want to 
accurately describe dwarf spheroidal galaxies within 1 kpc with SFDM haloes, then it suffices to use the base state of the SF. 

An important difference with the BEC model is that although eq.(\ref{eq:density}) provides comparable results to the simple BEC profile 
(eq.\ref{eq:density0}), eq.(\ref{eq:density}) comes from a model that provides a consistent description of galaxies with various sizes as 
opposed to the simple BEC model where all galaxies are assumed to reside in base state SFDM haloes.
If more data become available in the outer regions of DM dominated systems, specially in field dwarfs, it 
could be possible to test if their haloes are composed of more than just the usual Bose Einstein Condensates and also
the viability of the SFDM scenario as a DM alternative.

\section{Conclusions}\label{conclusions}

In this work we have shown that given the current data the scalar field dark matter gives a good description of the Milky Way dwarf 
spheroidal galaxies, we can model their DM haloes as ground state configurations of an ultra 
light scalar field. However, we found necessary to allow the existence of excited states in order to give a 
self consistent description of galaxies of different sizes and masses. The excited states together with the ground state form 
mixed state configurations where the ground or condensed state dominates the inner regions of 
DM dominated galaxies and the excited states account for the observations at large radii.
The first excited state do not change significantly the original fit in dwarf galaxies where most observations are located ($\sim$ 1 kpc),
but it allows the DM halo to extend to larger radius without supposing unrealistic anisotropy values. 
Thus, this natural extension removes the tension found by \citet{Diez2014}.

The question of stability of mixed configurations of scalar field similar to the ones we considered has been studied numerically in 
other works \citep{Bernal2010, Urena2010}.
In \citep{Bernal2010}, they present a stability study of multi-state configurations.
They show that stable configurations can be constructed when the number of particles in 
the first excited state is smaller than the number of particles in the ground state, which
is satisfied in all our fits, thus we expect our configurations to be long lived too.

Although there are some deviations with respect to the numerical solutions in the decaying behaviour of the density profile 
mainly due to the cut off radius that we impose taking the farthest observed value, we consider that our results will 
remain the same even for these numerical configurations as our analyses focused on the inner regions of galaxies which 
are independent on the particular asymptotic decaying behaviour.  

One additional remarkable feature of the SFDM haloes is that they predict constant central density profiles without the 
need of supernovae feedback in contrast to the divergent profiles characteristic of CDM simulations. 
Given that most dwarf haloes extend to $\approx 1$ $kpc$ and that we need only the ground state to reproduce the data within this radius, the DM mass enclosed in $r<1 kpc$ will always be smaller than CDM predictions, in fact, it might be that the mass reduction is just what is needed to account for the Too Big to Fail issue \citep{Boylan2011, gar14}, we will leave this discussion for a future work. 

As more accurate measurement in low mass and dwarf galaxies are obtained it will be possible to narrow the number of DM models 
that are consistent with the observations and get closer to understand the nature of the elusive dark matter, so far the 
ultra light scalar field dark matter model looks a promising and interesting candidate. 

\section*{Acknowledgments}
We thank Matthew Walker for providing access to the observational data.
The authors acknowledge to the General Coordination of Information and 
Communications Technologies (CGSTIC) at CINVESTAV for providing HPC 
resources on the Hybrid Cluster Supercomputer "Xiuhcoatl".
This work was partially supported by CONACyT M\'exico under grants CB-2009-01, no. 132400, CB-2011, no. 166212,  and I0101/131/07 C-234/07 of the Instituto Avanzado de Cosmologia (IAC) collaboration
(http://www.iac.edu.mx/). L. A. Martinez-Medina and V. H. Robles are supported by a CONACYT scholarship.

\label{lastpage}

\end{document}